\documentclass{aastex}          
\usepackage{spr-astr-addons}    

\begin{document}
%
\title{{\scriptsize \sf \vspace*{-44pt}Proceedings article for the HEDLA 2008 conference in St. Louis. Submitted for publication in\\[34pt]}
Comparison of Jupiter Interior Models Derived from First-Principles Simulations}

\shorttitle{Comparison of Jupiter Interior Models}
\shortauthors{Militzer and Hubbard}

\author{B. Militzer\altaffilmark{1}} 
\and 
\author{W. B. Hubbard\altaffilmark{2}}
\affil{}

\altaffiltext{1}{Departments of Earth and Planetary Science and of Astronomy, University of California, Berkeley, CA 94720, USA.}
\altaffiltext{2}{Lunar and Planetary Laboratory, The University of Arizona, Tucson, AZ 85721, USA.}

\begin{abstract}
Recently two groups used first-principles computer simulations to model
Jupiter's interior.  While both studies relied on the same simulation
technique, density functional molecular dynamics, the groups derived
very different conclusions. In particular estimates for the size of
Jupiter's core and the metallicity of its hydrogen-helium mantle
differed substantially. In this paper, we discuss the
differences of the approaches and give an explanation for the
differing conclusions.
\end{abstract}


The characterization of the interior structure of giant planets in our
solar system is crucial for identifying their formation mechanism and
for understanding the evolution of the solar system. Establishing the
history of our solar system will help interpreting the observed
similarities and differences between our and other solar systems. The
unexpected diversity among the over three hundred recently discovered
extra solar planets has challenged existing theories of planetary
formation and migration.

The planets in our solar system have been studied in great detail with
observations and space missions but many questions about their
interior structures have remained unanswered. Jupiter is predicted to
have a relatively small rocky core of between zero and seven Earth
masses~\citep{SC95,SG04}, which is surprising because similar theories
predicted between 10 and 25 Earth masses for the core in
Saturn~\citep{SG04}. This prediction for Jupiter has lent support to
core-accretion theories with comparatively small cores~\citep{Po96},
late-stage core erosion scenarios~\citep{Gu04}, or suggesting that
jovian planets are able to form directly from gases without a
triggering core~\citep{boss2007}.

The small core hypothesis for Jupiter has now been challenged in a
paper by Militzer, Hubbard, Vorberger, Tamblyn, and Bonev
(MHVTB)~\citep{MHVTB} who used first-principles computer simulations
of hydrogen-helium mixtures to compute the equation of state (EOS) in
the interior of Jupiter. This work predicts a large core of 14 -- 18
Earth masses for Jupiter, which is in line with estimates for Saturn
and suggests that both planets may have formed by core-accretion. The
paper further predicts small fraction of planetary ices in Jupiter's
envelope suggesting that the ices were incorporated into the core
during formation rather than accreted along with the gas envelope.
Jupiter is predicted to have an isentropic and fully convective
envelope that is of constant chemical composition. In order to match
the observed gravitational moment $J_4$, the authors suggest that
Jupiter may not rotate as a solid body and predicted the existence of
deep winds in the interior leading to differential rotation on
cylinders.

The first-principles simulations used in the MHVTB paper are a major
difference compared to chemical EOS models developed by Saumon,
Chabrier, and van Horn~\citep{SC92,SC95}. With first-principles
simulations one simulates a fully interacting quantum system of over a
hundred electrons and nuclei and therefore avoids a number of
approximations used in chemical models. In chemical models one for
example describes hydrogen as an ensemble of stable molecules, atoms,
free electrons, and protons and is then required to make additional
approximations to treat their interactions. These approximations are
dependent on the material under consideration and may also depend on the
temperature and pressure. First-principles methods describe the
interactions on a fundamental level. One is also required to make some
approximations to solve the many-body Schr\"odinger equation but those
are very different in nature, are not material specific and do not
depend on the $T$ and $P$. Therefore one expects EOSs derived from
first-principles to be more accurate than chemical models unless they
have been fit to experiments. However, no experimental EOS data are
available for the deep interiors of giant gas planets.

The MHVTB paper described significant differences between the
first-principles EOS and chemical models. However these were not
solely responsible for the different predictions for Jupiter's
interior. In a different paper, Nettelmann, Holst, Kietzmann, French,
Redmer and Blaschke (NHKFRB)~\cite{NHKFRB} also used first-principles
method to study Jupiter's interior but came to very different
conclusions. Very much in line with earlier models, they predict a
small core for Jupiter and a large amount of heavy elements in the
envelope.

In this paper, we will objectively discuss the differences between the
MHVTB and NHKFRB approaches in order to explain how such different
conclusions were derived, although we encourage the reader to compare
the two original papers also. The differences between the two papers
can be sorted into three categories: (1) differences in DFT-MD
simulations, (2) differences in the subsequent construction of
adiabats, and (3) different assumptions in the models for Jupiter's
interior. We will go through these differences in the following three
sections and demonstrate that the main difference arises from point
(3).

\section{Comparison of Simulation Parameters}

In this section we discuss the differences in the first-principles
simulations performed by the two groups. Although there are
differences in the level of accuracy, they are unlikely to be the main
reason for the differences in the Jupiter models. Both groups used
density functional molecular dynamics (DFT-MD) simulations and should
derive identical equations of state for hydrogen, helium, and their
mixtures. However, computational details are important for accuracy of
the derived EOS.

The NHKFRB group used exclusively VASP code while MHVTB used CPMD as
well as the VASP code. MHVTB verified that both codes yield identical
results when run accurately.

The MHVTB group performed simulations of hy\-dro\-gen-helium mixtures
using a mixing ratio that closely resembles Jupiter's composition within
the limitations of a finite simulation cell. The NHKFRB instead relied
on the ideal mixing approximation using EOSs of pure hydrogen and
helium. This approximation neglects all interactions between hydrogen
and helium. In Refs.~\citep{Vo07,Vo07b}, it was shown that this
interaction leads to significant corrections in the derived pressures
and energy but, more importantly, the presence of helium increases
that stability of the hydrogen molecules for given $T$ and $P$. These
interaction effects also have an impact on the derived adiabat that
will be discussed in the next section. However, linear mixing
approximation does not affect the predicted core mass very much. In
fact, the MHVTB group used their own set of simulations for pure
hydrogen and helium~\citep{Mi08} in a separate Jupiter model and
derived a core mass that differed by only 2 Earth masses.  MHVTB use
this independent calculation to estimate the uncertainty of the
predictions.

The NHKFRB group performed DFT-MD simulations with 500-2000
time steps. MHVTB performed simulations for 2 picoseconds with 5000
time steps. Longer simulations lead to more accurate averages for
thermodynamic variables such as pressure and internal energy.

Concerning the size of the simulation, the NHKFRB paper states that
simulations with between 32 and 162 atoms for hydrogen, helium, and
water were performed. A conservative estimate of the precision of the
EOS 5\% is given. Conference presentations by this group showed
results with 64 hydrogen atoms and 32 helium atoms (64 electrons). The
MHVTB group used simulations with 128 electrons throughout, which are
more accurate.

The NHKFRB used $(\frac{1}{4},\frac{1}{4},\frac{1}{4})$ Baldereschi
point to sample the Brillouin zone for all simulations. The MHVTB
group analyzed the $k$-point dependence using simulation with up to
4x4x4 k-points and then performed simulation with 2x2x2 in the
metallic regime and using the $\Gamma$ point at lower densities. It
was shown that $\Gamma$ point only simulation overestimate the
pressure near Jupiter's core-mantle boundary by 1.6\%.

\section{Derivation of the Adiabats}

Convection dominates the heat transport in giant gas planets. This
leads to an adiabatic temperature-pressure profile for the planet's
interior with the exception of a small low pressure region near the
surface where radiative heat transport takes over. It is not clear if
the cores of giant planets themselves are convective but the following
planetary models are insensitive to the temperature profile in the
core. On the other hand, the temperature profile in the
hydrogen-helium rich envelope and the required derivation of adiabats
are very important to characterize the interiors of giant
planets. However, neither Monte Carlo nor molecular dynamics methods
can compute the entropy directly because both techniques save orders
of magnitude of computer time by generating only a representative
sample of configurations instead of integrating over the whole
configuration space that would be needed to derive the entropy.

One typically derives the entropy by thermodynamic integration from a
known reference state. This can be very computationally demanding and
is also not needed to map out planetary interiors. The absolute value
of the entropy is not important as long as one is able to construct
$(T,P)$ curves of constant entropy. This can be achieved by using the
pressure and the internal energy from first-principles simulations at
different $(T,V)$ conditions. Using Maxwell's relations, one finds,
\begin{equation}
\left. \frac{\partial T}{\partial V}\right|_S = - \frac
{ \left. \frac{\partial S}{\partial V}\right|_T }
{ \left. \frac{\partial S}{\partial T}\right|_V } 
= - T \frac
{ \left. \frac{\partial P}{\partial T}\right|_V }
{ \left. \frac{\partial E}{\partial T}\right|_V } \;.
\label{entropy}
\end{equation}
By solving this ordinary differential equation, $(V,T)$-adiabats can
be constructed as long as a sufficiently dense mesh of high-quality
EOS points are available to make the required interpolation and
differentiation of $E$ and $P$ with respect to temperature
satisfactorily accurate.

One drawback of formula~(\ref{entropy}) is that it is not necessarily
thermodynamically consistent if pressures and internal energies are
interpolated separately. This is the primary reason why we developed a
fit for the free energy~\citep{Mi08} that is thermodynamically
consistent by construction. The free energy, $F(V,T)$, is fit in such
a way that first-principles pressures and internal energies are
reproduced by,
\begin{equation}
P =   -\left. \frac{\partial F}{\partial V}\right|_T \;\;\;\;\;\;{\rm and}\;\;\;\;\;\;
E = F -T \left. \frac{\partial F}{\partial T}\right|_V\;.
\end{equation}

The details of this method are described in Ref.~\citep{Mi08} where a
helium EOS derived from first-principles simulations is presented for
a large temperature and pressure interval.

We apply this fit across the insulator-to-metal transition in fluid
hydrogen-helium mixtures. According to predictions from the best
simulation methods currently available, quantum Monte
Carlo~\citep{Delaney06} and DFT-MD~\citep{Vo07} this transition is
expected to occur gradually. However, for pure hydrogen, the
dissociation transitions gives raise to a region of negative $\partial
P / \partial T|_V$~\citep{Vo07}, which leads to a negative Gr\"uneisen
parameter and might introduce a barrier into Jupiter's
convection. However, the MHVTB work demonstrated that in a
hydrogen-helium mixture the region of $\partial P / \partial T|_V < 0$
is shifted to lower temperatures than occur in Jupiter. Significant
effort went into analyzing this transition within the DFT-MD
method. The conclusion is that Jupiter's interior is fully convective
and this is main reason for constructing a two-layer model.

The dissociation transition leads to a region where $\partial P /
\partial T|_V$ is small but positive. Following Eq. (\ref{entropy}), 
the temperature along the adiabat raise very little in this
region. As a result, the predicted temperatures for the metallic
regime reaching all the way down to core-mantle boundary are
significantly lower than predicted by the NHKFRB model. There is no
information in the NHKFRB paper on how the entropy was derived. At
this point, one cannot verify whether Eq.~(\ref{entropy}) is
fulfilled. The NHKFRB adiabats agree well with adiabat from the SCvH
EOS model for lowest and highest pressures in Jupiter's envelope.

The MHVTB adiabats, on the other hand agree with SCvH only in the low
density limit. At the highest pressures at the core-mantle boundary,
one still finds differences in the predicted temperatures much in the
same way that differences in pressure are observed. At Jupiter's
core-mantle boundary the pressures are not yet high enough so that
the hydrogen-helium mixtures would be mostly ideal. Therefore, in the
absence of experimental data, differences between {\it ab initio} and
chemical models are expected.

\section{Comparison of Jupiter Interior Models}

The main difference between the two Jupiter's model under
consideration arise from the modeling assumptions. The MHVTB group
derived a new type of Jupiter model that has only {\it two layers}: a
dense core and a completely isentropic (consistent with full
convection and no phase transitions) mantle composed of mostly
hydrogen and helium. Because there is no freedom in this type of model
to redistribute mass by invoking chemical discontinuities in the
mantle, the group could only match the gravity moment $J_4$ and all
other constraints by invoking differential rotation with deep winds in
Jupiter's envelope instead of the conventional solid-body rotation
with minor surface winds. Substantial differential rotation in
Jupiter, potentially detectable by the forthcoming Juno orbiter, was
one of main predictions of the MHVTB paper.

The NHKFRB group used a {\it three layer model} that is much closer to
previous models by~\cite{SG04}. Besides a solid rock core, NHKFRB
model assumes the mantle to be composed of two layers with differing
composition.  The flexibility to distribute helium and heavier
elements unevenly in the two mantle layers and the associated
redistribution of mass allows NHKFRB group to match $J_4$ without
additional assumptions such as differential rotation. {\it This is
main difference between the two models.} The redistribution of mass in
the NHKFRB model to match $J_4$ also reduces the predicted core mass
and leads to a larger amount of heavier elements in the envelope.

In the NHKFRB model, the inner layer of Jupiter is helium rich
(24.5\%) and contains metallic hydrogen while the outer layer is
helium poor (23.3\%) and contains molecular hydrogen. However, the
NHKFRB model also predicts a very large difference in the
concentration of heavier elements: 2.1\% in the outer and 16.6\% in
inner layer. An explanation how such large concentration differences
can arise during Jupiter's evolution remains to be given. In this
regard, the MHVTB is much simpler. It assumes a fully convective
mantle of constant chemical composition.


NHKFRB used the standard theory of figures to derive Jupiter's gravity
field. In contrast, MHVTB used two independent approaches to derive
the gravity field.  The first approach used the theory of figures,
while the second approach used the self-consistent-field method
incorporating arbitrary differential
rotation~\citep{Hubbard1975,Hubbard1982} to monitor numerical errors
in the gravity field calculations and to confirm that the $J_4$
mismatch for solid-body rotation is not a numerical artifact.  Because
NHKFRB used a standard solution for the uniformly-rotating polytrope
of index one to test their code, it seems unlikely that their gravity
calculations are affected by errors either.  However, as a minor
point, NHKFRB fitted their models to an older value for $J_4$ with
larger error bars instead of using~\citep{citation24}.

Summarizing one can say that MHVTB predicted a large core in Jupiter
of 14 -- 18 Earth masses using a new two-layer model with a fully
convective envelope. The NHKFRB work predicted a much smaller core
based on a three-layer model that is very similar to earlier Jupiter
models. The crucial difference lies in the treatment of the
molecular-to-metallic transition in dense fluid hydrogen and more work
is needed to characterize this transition with different experimental
and theoretical techniques. MHVTB analyzed this transition within
density functional theory and concluded that this transition is
continuous leading to an Jupiter envelope of constant chemical
composition. Instead of analyzing this transition, NHKFRB follow
previous models and made the additional assumption of two chemically
different mantle layers. Besides the discussed accuracy differences in
the computed EOS, we attribute this extra assumption to be the
primary reason with the difference in the predicted core masses for
Jupiter.

\acknowledgments
We thank J. Vorberger and D. Stevenson for comments.


%

%

\end{document}